\documentclass[aps,twocolumn,showpacs,superscriptaddress
]{revtex4}
\usepackage{epsf}
\begin{document}
\title{Multicanonical Study of Coarse-Grained Off-Lattice Models for Folding Heteropolymers}
\author{Michael Bachmann}
\email[E-mail: ]{Michael.Bachmann@itp.uni-leipzig.de}
\affiliation{Institut f\"ur Theoretische Physik, Universit\"at Leipzig,
Augustusplatz 10/11, D-04109 Leipzig, Germany}
\author{Handan Ark{\i}n}
\email[E-mail: ]{Handan.Arkin@itp.uni-leipzig.de}
\affiliation{Institut f\"ur Theoretische Physik, Universit\"at Leipzig,
Augustusplatz 10/11, D-04109 Leipzig, Germany}
\affiliation{Department of Physics Engineering, Hacettepe University,
06532 Ankara, Turkey}
\author{Wolfhard Janke}
\email[E-mail: ]{Wolfhard.Janke@itp.uni-leipzig.de}
\homepage[\\ Homepage: ]{http://www.physik.uni-leipzig.de/CQT}
\affiliation{Institut f\"ur Theoretische Physik, Universit\"at Leipzig,
Augustusplatz 10/11, D-04109 Leipzig, Germany}
\begin{abstract}
We have performed multicanonical simulations of hydrophobic--hydrophilic heteropolymers with two simple effective,
coarse-grained off-lattice models to study the influence of specific interactions in the models
on conformational transitions of selected sequences with 20 monomers. Another aspect 
of the investigation was the comparison with the purely hydrophobic homopolymer
and the study of general conformational properties induced by the ``disorder''
in the sequence of a heteropolymer.  
Furthermore, we applied an optimization algorithm to sequences with up to 55 monomers
and compared the global-energy 
minimum found with lowest-energy states identified within the multicanonical simulation.
This was used to find out how reliable the multicanonical method samples the free-energy landscape,
in particular for low temperatures. 
\end{abstract}
\maketitle
\section{Introduction}
\label{intro}
The understanding of protein folding is one of the most challenging
objectives in biochemically motivated research. Although the physical
principles are known, the complexity of proteins as being macromolecules
consisting of numerous atoms, the influence of quantum chemical details 
on long-range interactions as well as the role of the solvent, etc.,  
makes an accurate analysis of the folding process of realistic
proteins extremely difficult. Therefore, one of the most important questions 
in this field is how much detailed information can be neglected to establish
effective, coarse-grained models yielding reasonable, at least qualitative, results that
allow for, e.g., a more global view on the relationship between the
sequence of amino acid residues and the existence of a global, funnel-like
energy minimum in a rugged free-energy landscape~\cite{onuchic1}. 

Within the past two decades much work has been devoted to introduce minimalistic
models based on general principles that are believed to primarily control
the structure formation of proteins. 
One of the most prominent examples
is the HP model of lattice proteins~\cite{dill1} which has been exhaustively investigated
without revealing all secrets, despite its simplicity. In this model, only two types
of monomers are considered, with hydrophobic ($H$) and polar ($P$) character. Chains on the lattice are 
self-avoiding to account for the excluded volume. The only explicit interaction is between
non-adjacent but next-neighbored hydrophobic monomers. This interaction of hydrophobic
contacts is attractive to force
the formation of a compact hydrophobic core which is screened from the (hypothetic)
aqueous environment by the polar residues. Statistical mechanics simulations of this 
model are still subject of studies 
requiring the application of sophisticated algorithms~\cite{iba1,hsu1,bj1}. 

A manifest
off-lattice generalisation of the HP model is the AB model~\cite{ab1}, where the hydrophobic monomers are
labelled by $A$ and the polar or hydrophilic ones by $B$. The contact interaction is replaced by
a distance-dependent Lennard-Jones type of potential accounting for 
short-range excluded volume repulsion and long-range interaction, the latter being attractive for $AA$ and
$BB$ pairs and repulsive for $AB$ pairs of monomers. An additional interaction accounts for the bending
energy of any pair of successive bonds. This model was first applied in two dimensions~\cite{ab1} and 
generalized to three-dimensional AB proteins~\cite{irb1,irb2}, 
partially with modifications taking implicitly into account additional 
torsional energy contributions of each bond. 

More knowledge-based coarse-grained models are often of G\={o} type, where, in the simplest lattice 
formulation, the energy is 
proportional to the negative number of native contacts. These models usually require the knowledge of the 
native conformation and serve, e.g., as models for studies of folding pathways~\cite{pande1,weikl1} and native 
topology~\cite{clementi1, koga1}. There is also growing interest in modeling the folding behavior of
single-domain proteins with simplified models as many of them show up a simple two-state kinetics~\cite{chan1}
without intermediary states that would slow down the folding dynamics (``traps''). It seems,
however, that native-centric models such as those of G\={o} type require modifications for a
qualitatively correct description of this sharp folding transition~\cite{kaya1,kaya2}.  

In this paper we report studies of thermodynamic and ground-state properties of AB sequences known 
from the literature~\cite{irb1,hsu2,liang1}
for two representations of the AB model~\cite{ab1,irb1} which are described in Sec.~\ref{models}.
While, compared to all-atom formulations, the interactions in these coarse-grained models 
are greatly simplified and hence can be computed much faster, they still preserve a complicated 
rugged free-energy landscape, where naive simulations would easily get trapped. 
In order to accurately resolve the low-temperature behavior we, therefore, applied a multicanonical Monte Carlo
algorithm~\cite{muca1,muca2}
with an appropriate update mechanism. For simulations of systems with complex free energy landscapes, multicanonical sampling 
has become very popular and its application in protein simulations has a long tradition~\cite{hansmann0}. 
The ground-state
search was achieved by means of the energy landscape paving minimizer (ELP)~\cite{hansmann1}. 
Section~\ref{secmeth} is devoted to the description of these methods.
We present the results for the global energy minima and the thermodynamic quantities in Sec.~\ref{results}. 
The paper is concluded by the summary in Sec.~\ref{summary}.
\section{Effective Off-Lattice Models}
\label{models}
We investigated two effective off-lattice models of AB type for heteropolymers
with $N$ monomers. The first one is the original
AB model as proposed in Ref.~\cite{ab1} with the energy function
\begin{eqnarray}
\label{stillinger}
E_I &=& \frac{1}{4}\sum\limits_{k=1}^{N-2}(1-\cos \vartheta_k)+\nonumber\\
&&\hspace{7mm} 4\sum\limits_{i=1}^{N-2}\sum\limits_{j=i+2}^N\left(\frac{1}{r_{ij}^{12}}
-\frac{C_I(\sigma_i,\sigma_j)}{r_{ij}^6} \right),
\end{eqnarray}
where the first sum runs over the $(N-2)$ angles $0\le \vartheta_k\le \pi$ of successive bond vectors. 
This term is the bending energy and the coupling is ``ferromagnetic'', i.e., it costs
energy to bend the chain. The second term partially competes with the bending barrier by a 
potential of Lennard-Jones type depending on the distance between monomers being non-adjacent 
along the chain. It also accounts for the influence of the AB sequence ($\sigma_i=A$ for hydrophobic
and $\sigma_i=B$ for hydrophilic monomers) on the energy of a conformation
as its long-range behavior is attractive for pairs of like monomers and repulsive for $AB$ pairs
of monomers:
\begin{equation}
\label{stillC}
C_I(\sigma_i,\sigma_j)=\left\{\begin{array}{cl}
+1, & \hspace{7mm} \sigma_i,\sigma_j=A,\\
+1/2, & \hspace{7mm} \sigma_i,\sigma_j=B,\\
-1/2,  & \hspace{7mm} \sigma_i\neq \sigma_j.\\
\end{array} \right.    
\end{equation}
We will refer to this model as AB model I throughout the paper. 

The other model we have studied
has been introduced in Ref.~\cite{irb1}  and is a variant of AB model I in that it also consists 
of angular and distance-dependent energy terms, but with some substantial modifications. 
In the following, we denote it as AB model II. The energy is given by
\begin{eqnarray}
\label{irbaeck}
E_{II}&=&-\kappa_1\sum\limits_{k=1}^{N-2}{\bf b}_k\cdot{\bf b}_{k+1}-
\kappa_2\sum\limits_{k=1}^{N-3}{\bf b}_k\cdot{\bf b}_{k+2}+\nonumber\\
&&\hspace{3mm}4\sum\limits_{i=1}^{N-2}\sum\limits_{j=i+2}^NC_{II}(\sigma_i,\sigma_j)\left(\frac{1}{r_{ij}^{12}}
-\frac{1}{r_{ij}^6} \right),
\end{eqnarray}
where ${\bf b}_k$ is the bond vector between the monomers $k$ and $k+1$ with length unity. In Ref.~\cite{irb1}
different values for the parameter set ($\kappa_1$, $\kappa_2$) were tested and finally set to ($-1$, $0.5$) 
as this choice led to distributions for the angles between bond vectors ${\bf b}_k$ and ${\bf b}_{k+1}$
as well as the torsion angles between the surface vectors ${\bf b}_k\times{\bf b}_{k+1}$ 
and ${\bf b}_{k+1}\times{\bf b}_{k+2}$ that agreed best with distributions obtained for selected 
functional proteins. Since ${\bf b}_k\cdot{\bf b}_{k+1}=\cos \vartheta_k$, the choice $\kappa_1=-1$ makes
the coupling between successive bonds ``antiferromagnetic'' or ``antibending'' contrary to what was
chosen in Eq.~(\ref{stillinger}) for AB model I. The second term in Eq.~(\ref{irbaeck}) takes torsional 
interactions into account without being an energy associated with the pure torsional barriers in the usual 
sense. The third term contains now a pure Lennard-Jones potential, where the $1/r_{ij}^6$ long-range interaction
is attractive whatever types of monomers interact. The monomer-specific prefactor $C_{II}(\sigma_i,\sigma_j)$
only controls the depth of the Lennard-Jones valley:
\begin{equation}
\label{irbC}
C_{II}(\sigma_i,\sigma_j)=\left\{\begin{array}{cl}
+1, & \hspace{7mm} \sigma_i,\sigma_j=A,\\
+1/2, & \hspace{7mm} \sigma_i,\sigma_j=B\quad \mbox{or}\quad \sigma_i\neq \sigma_j.
\end{array} \right.    
\end{equation}    
For technical reasons, we have introduced in both models a cut-off $r_{ij}=0.5$ for the Lennard-Jones potentials
below which the potential is hard-core repulsive (i.e., the potential is infinite).  
For both models only a few results are given in the literature. For the first model these are estimates for the
global energy minima of certain AB sequences~\cite{hsu2}, while for AB model II primarily thermodynamic quantities
were determined~\cite{irb1}. We have performed ELP optimizations~\cite{hansmann1}
in order
to find deeper-lying energy minima than the values quoted and, in particular, multicanonical 
sampling~\cite{muca1,muca2} for
enabling us
to focus on thermodynamic properties of the sequences given in Ref.~\cite{irb1}.  
\section{Methods}
\label{secmeth}
In this section we describe the computational sampling methods and the update
procedure we applied to obtain results for off-lattice AB heteropolymers.
\subsection{Energy-landscape paving optimization procedure}
\label{secelp}
In order to locate global energy minima of a complex system, it is often useful to apply specially biased
algorithms that only serve this purpose. We used the energy landscape paving (ELP) minimizer~\cite{hansmann1}
to find global energy minima of the sequences under consideration. The ELP minimization
is a Monte Carlo optimization method,
where the energy landscape is locally flattened. This means that if a state ${\bf x}$ with energy 
$E({\bf x})$ is hit, the energy is increased by a ``penalty'' which itself depends
on the histogram of any suitably chosen order parameter. The simplest choice is the energy distribution
$H(E)$ such that we define $\tilde{E}_t = E_t+f(H(E_t))$. Thus, the Boltzmann
probability $\exp(-\tilde{E}_t/k_BT)$ for a Metropolis update, 
where $k_BT$ is the thermal energy at the temperature $T$,
becomes a function of ``time'' $t$.
The advantage of this method is that local energy minima are filled up and the likelihood of touching 
again recently visited regions decreases. This method has successfully proved to be applicable to
find global energy minima in rough energy landscapes of proteins~\cite{hansmann1,ha1}.
Of course, as a consequence of the biased sampling, stochastic methods along the line of ELP violate
detailed balance and it is therefore inappropriate to apply those for uncovering thermodynamic
properties of a statistical system.
\begin{table*}[t]
\caption{\label{tabGEM_A} The four Fibonacci sequences, for which we analyzed the ground states in both models.}
\begin{tabular}{ll}\hline\hline
$N$ & \multicolumn{1}{c}{sequence} \\ \hline
13 & AB$_2$AB$_2$ABAB$_2$AB \\
21 & BABAB$_2$ABAB$_2$AB$_2$ABAB$_2$AB \\
34 & AB$_2$AB$_2$ABAB$_2$AB$_2$ABAB$_2$ABAB$_2$AB$_2$ABAB$_2$AB \\
55 & BABAB$_2$ABAB$_2$AB$_2$ABAB$_2$ABAB$_2$AB$_2$ABAB$_2$AB$_2$ABAB$_2$ABAB$_2$AB$_2$ABAB$_2$AB \\ \hline \hline
\end{tabular}
\end{table*}
\begin{table*}[t]
\caption{\label{tabGEM_B} Estimates for global energy minima obtained with multicanonical 
(MUCA) sampling and ELP minimization~\cite{hansmann1} for the Fibonacci sequences of Table~\ref{tabGEM_A}
using both models. 
The values for AB model I are compared with results quoted in Ref.~\cite{hsu2}
employing off-lattice PERM and after subsequent conjugate-gradient minimization (PERM+).
Minimum energies found with MUCA and ELP for AB model II are compared with the 
lowest energies listed in Ref.~\cite{liang1}, obtained with annealing contour 
Monte Carlo (ACMC) and improved by Metropolis quenching (ACMC+). 
}
\begin{tabular}{l|cc|cc|cc|cc}\hline\hline
& \multicolumn{4}{c|}{AB model I} & \multicolumn{4}{c}{AB model II}\\
$N$ & $E_{\rm min}^{\rm MUCA}$ & $E_{\rm min}^{\rm ELP}$ & 
$E_{\rm min}^{\rm PERM}$~\cite{hsu2} & $E_{\rm min}^{\rm PERM+}$~\cite{hsu2} &
$E_{\rm min}^{\rm MUCA}$ & $E_{\rm min}^{\rm ELP}$ & 
$E_{\rm min}^{\rm ACMC}$~\cite{liang1} & $E_{\rm min}^{\rm ACMC+}$~\cite{liang1}
\\ \hline
13 & $-4.967$ & $-4.967$ & $-3.973$ & $-4.962$ & $-26.496$ & $-26.498$ & $-26.363$ & $-26.507$ \\ 
21 & $-12.296$ & $-12.316$ & $-7.686$ & $-11.524$ & $-52.915$ & $-52.917$ & $-50.860$ & $-51.718$ \\  
34 & $-25.321$ & $-25.476$ & $-12.860$ & $-21.568$ & $-97.273$ & $-97.261$ & $-92.746$ & $-94.043$\\
55 & $-41.502$ & $-42.428$ & $-20.107$ & $-32.884$ & $-169.654$ & $-172.696$ & $-149.481$ & $-154.505$\\ \hline \hline
\end{tabular}
\end{table*}
\subsection{Multicanonical method}
\label{secmuca}
To obtain statistical results we applied a multicanonical Monte Carlo algorithm, where the 
energy distribution is flattened artificially allowing, in principle, for a random walk of successive states
in energy space. This flattening is controllable and therefore reproducible. For this purpose, the
Boltzmann probability is multiplied by a weight factor $W(E)$, which in our case is a function of the 
energy. Then the probability for a state with energy $E$ reads $p_M(E)=\exp(-E/k_BT)W(E)$. In order
to obtain a multicanonical or ``flat'' distribution, the initially unknown weight function $W(E)$ has
to be chosen accordingly what is done iteratively: In the beginning, the weights $W^{(0)}(E)$ are set to 
unity for all energies letting the first run be a usual Metropolis simulation which yields an
estimate $H^{(0)}(E)$ for the canonical distribution. The histogram is used to determine the 
next guess for the weights, the simplest update is to calculate $W^{(1)}(E)=H^{(0)}(E)/W^{(0)}(E)$.
Then the next run is performed with probabilities $p_M^{(1)}(E)=\exp(-E/k_BT)W^{(1)}(E)$ of states
with energy $E$. The iterative procedure is continued until 
the weights are appropriate in a way that the histogram is ``flat''. The introduction
of a flatness criterion, for instance, that the fluctuations around the average value of the histogram 
are less than 20\%, is useful, but not necessary, if the number of iterations is very large.
After having determined accurate weights $W(E)$, they are kept fixed
and following some thermalization sweeps a long production run is performed, where statistical
quantities $O$ are obtained multicanonically, 
$\langle O\rangle_M=\sum_{\{\bf x\}}p_M(E(\{{\bf x}\}))O(\{{\bf x}\})/Z_M$ with the multicanonical
partition function $Z_M=\sum_{\{\bf x\}}p_M(E(\{{\bf x}\}))$. The canonical statistics 
is obtained by reweighting the multicanonical to the canonical distribution, i.e., 
mean values are computed as $\langle O\rangle=\langle OW^{-1} \rangle_M/\langle W^{-1} \rangle_M$.   
    
For the determination of the multicanonical weights we performed $200$ iterations with 
at least $10^5$ sweeps each. The histograms obtained in the iteration runs were accumulated 
error-weighted~\cite{muca2} such that the estimation of the weights was based on increasing statistics
and not only on the relatively small number of sweeps per run. In the production period, $5\times 10^7$ 
sweeps were generated to have reasonable statistics for estimating the thermodynamic quantities.  
Statistical errors are estimated with the standard Jackknife technique~\cite{jack1,jack2}.
\subsection{Spherical update mechanism}
\label{secupdate}
\begin{figure}
\centerline{\epsfxsize=8.8cm \epsfbox{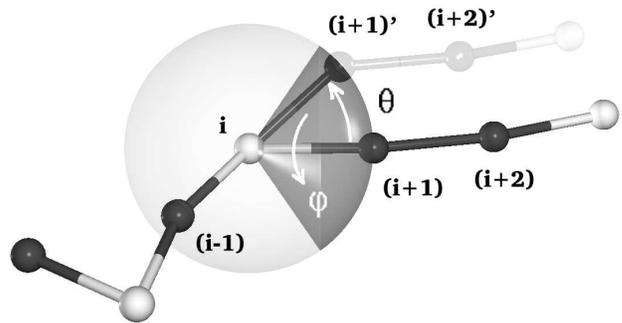}}
\caption{\label{figupdate} Spherical update of the bond vector between the $i$th and
$(i+1)$th monomer.  } 
\end{figure}
For updating a conformation we use the procedure displayed in Fig.~\ref{figupdate}. 
Since the length of the bonds is fixed ($|{\bf b}_k|=1$, $k=1,\ldots,N-1$), the $(i+1)$th monomer
lies on the surface of a sphere with radius unity around the $i$th monomer.
Therefore, spherical coordinates are the natural choice for calculating the new
position of the $(i+1)$th monomer on this sphere. For the reason of efficiency,
we do not select any point on the sphere but restrict the choice to a
spherical cap with maximum opening angle $2\theta_{\max}$ (the dark area in
Fig.~\ref{figupdate}). Thus, to change the position of the $(i+1)$th monomer to
$(i+1)'$, we select the angles $\theta$ and $\varphi$ randomly from the 
respective intervals $\cos \theta_{\max} \le \cos \theta \le 1$ and $0\le\varphi \le 2\pi$,
which ensure a uniform distribution of the $(i+1)$th monomer positions on the associated 
spherical cap.
After updating the position of the
$(i+1)$th monomer, the following monomers in the chain are simply translated 
according to the corresponding bond vectors which remain unchanged in this type
of update. Only the bond vector between the $i$th and the $(i+1)$th monomers is
rotated, all others keep their direction. This is similar to single spin updates
in local-update Monte Carlo simulations of the classical Heisenberg model with the 
difference that in addition to local energy changes long-range interactions of the monomers, changing their relative position 
to each other, have to be computed anew after the update.
In our simulations of the AB models we used a very small opening angle, $\cos\theta_{\max}=0.99$,
in order to be able to sample also very narrow and deep valleys in the landscape
of angles.
\section{Results}
\label{results}
We applied the multicanonical algorithm primarily to study thermodynamic
properties, e.g., the ``phase'' behavior of off-lattice heteropolymers.
Before discussing these results, however, we analyze the capability of multicanonical
sampling to find lowest-energy conformations, in particular the native fold, 
because the identification of these structures is not only interesting as a
by-product of the simulation. Rather, since they dominate the low-temperature
behavior, it is necessary that they are generated frequently in the multicanonical
sampling. 
\subsection{Search for global energy minima}
\label{GEM}
\begin{table}[t]
\caption{\label{tabRMSD} Root mean square deviations rmsd and overlap $Q$ of lowest-energy conformations
found with multicanonical sampling and with the ELP optimizer for the Fibonacci sequences of length $N$
given in Table~\ref{tabGEM_A}.}
\begin{tabular}{l|cc|cc}\hline\hline
& \multicolumn{2}{c|}{AB model I} & \multicolumn{2}{c}{AB model II}\\ 
$N$ & rmsd & $Q$ & rmsd & $Q$\\ \hline
13 & $0.015$ & $0.994$ & $0.006$ & $0.998$ \\
21 & $0.025$ & $0.992$ & $0.009$ & $0.997$ \\
34 & $0.162$ & $0.979$ & $1.412$ & $0.840$ \\
55 & $2.271$ & $0.766$ & $1.904$ & $0.857$ \\ \hline \hline
\end{tabular}
\end{table}
In order to be able to investigate the low-temperature behavior of AB off-lattice heteropolymers,
we first have to assess the capabilities of the multicanonical method to sample lowest-energy
conformations and, in particular, to approach closely the global energy minimum conformation.
In Table~\ref{tabGEM_A} we list the Fibonacci sequences~\cite{ab1} that 
have already been studied in Ref.~\cite{hsu2}
by means of an off-lattice variant of the improved version~\cite{hsu1} of the celebrated chain-growth 
algorithm with population control, PERM~\cite{grass1}. In Ref.~\cite{hsu2}, first estimates for
the putative ground-state energies of the Fibonacci sequences with 13 to 55 monomers were given
for AB model I~\cite{ab1} in three dimensions. We compare these results with the respective
lowest energies found in our multicanonical simulations and with the results obtained with
the minimization algorithm ELP~\cite{hansmann1}. It turns out that the ground-state
energies found by multicanonical sampling agree well with what comes out by the ELP minimizer,
cf.\ Table~\ref{tabGEM_B}. Another interesting result is that our estimates for the ground-state
energies lie significantly below the energies quoted in Ref.~\cite{hsu2}, 
obtained with the off-lattice PERM
variant, and our values are even lower than the energies obtained by PERM and subsequent 
conjugate-gradient minimization in the attraction basin. 

We have also performed this test with model II~\cite{irb1} and we
can compare our results with minimum energies listed in Ref.~\cite{liang1}, where the so-called 
annealing contour Monte Carlo (ACMC) method was applied to these Fibonacci sequences. From 
Table~\ref{tabGEM_B} we see that we find with MUCA and ELP runs lower energies for the sequences with
21, 34, and 55 monomers, while the results for the 13mer are comparable.

Note that the multicanonical algorithm is not tuned to give good results in the low-energy sector
only. For all sequences studied in this work, the same algorithm also yielded the thermodynamic results to be
discussed in the following section. 
\begin{figure}[t]
\parbox{4.3cm}{
\centerline{\epsfxsize=4.2cm \epsfbox{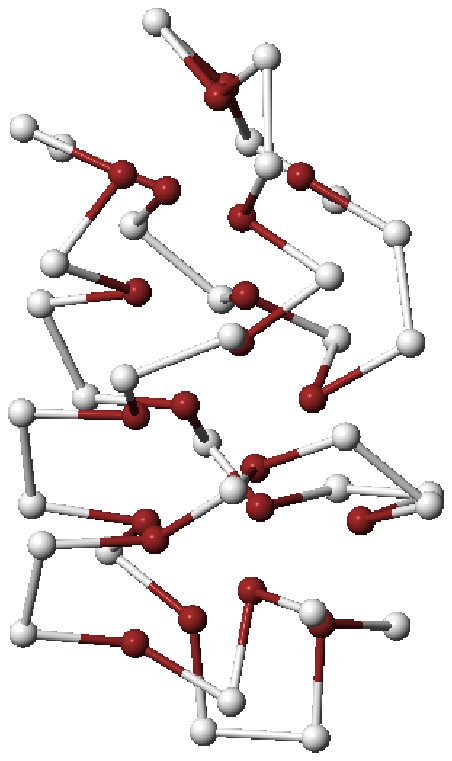}}}\hfill
\parbox{4.3cm}{
\centerline{\epsfxsize=4.2cm \epsfbox{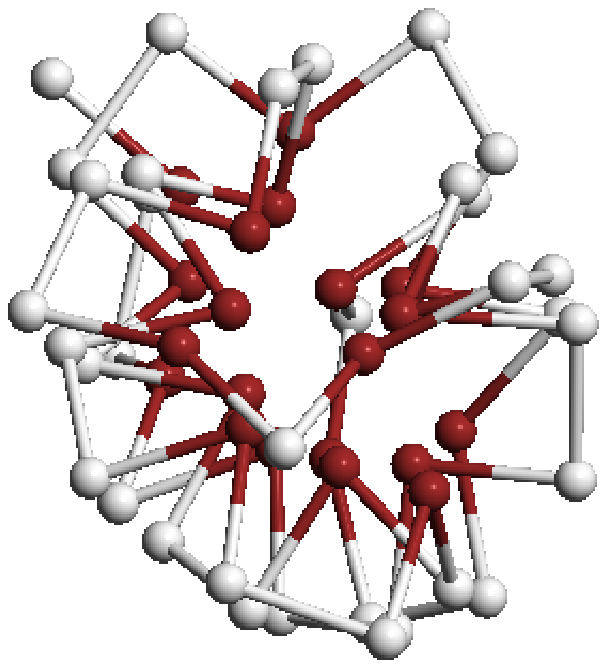}}}
\caption{\label{min55elp} 
Side (left) and top view (right) of the global energy minimum conformation
of the 55mer (AB model I) found with the ELP minimization algorithm (dark spheres: hydrophobic
monomers -- A, light spheres: hydrophilic -- B).
}
\end{figure}

In order to check the structural similarities of the lowest-energy conformations obtained
with multicanonical sampling and those from the ELP optimization runs, 
${\bf X}^{\rm MUCA,ELP}=({\bf x}^{\rm MUCA,ELP}_1,{\bf x}^{\rm MUCA,ELP}_2,\ldots, 
{\bf x}^{\rm MUCA,ELP}_N)$, we calculated the rmsd (root mean square deviation) of the respective pairs,
\begin{equation}
\label{rmsd}
{\rm rmsd} = \min \sqrt{\frac{1}{N}\sum\limits_{i=1}^N
|\tilde{\bf x}^{\rm MUCA}_i-\tilde{\bf x}^{\rm ELP}_i|^2}.
\end{equation}
Here, $\tilde{\bf x}^{\rm MUCA,ELP}_i={\bf x}^{\rm MUCA,ELP}_i-{\bf x}^{\rm MUCA,ELP}_0$ 
denotes the position 
with respect to the respective centers of mass ${\bf x}_0=\sum_{j=1}^N{\bf x}_j/N$ of the $i$th monomer
of the lowest-energy conformations
found with multicanonical sampling and ELP minimization, respectively. Obviously, the
rmsd is zero for exactly coinciding conformations and the larger the value the worse
the coincidence. The minimization of the sum in Eq.~(\ref{rmsd}) is performed with
respect to a global relative rotation
of the two conformations in order
to find the best match.  
For the explicit calculation we
used the exact quaternion-based optimization procedure described in Ref.~\cite{kearsley1}. 

As an alternative, we introduce here another parameter that enables us to compare conformations.
Instead of performing comparisons of positions it is much simpler and less time-consuming
to calculate the so-called overlap between two conformations by comparing their bond and torsion
angles. As an extension of the torsion-angle based variant~\cite{okamoto1,okamoto2},
we define the more general overlap parameter
\begin{equation}
\label{overlap}
Q({\bf X},{\bf X'})=\frac{N_t+N_b-d({\bf X},{\bf X'})}{N_t+N_b},
\end{equation}
where (with $N_t=N-3$ and $N_b=N-2$ being the numbers of torsional angles $\Phi_i$ and 
bond angles $\Theta_i=\pi-\vartheta_i$, respectively)
\begin{eqnarray*}
\label{dparam}
d({\bf X},{\bf X'})&=&
\frac{1}{\pi}\left(\sum\limits_{i=1}^{N_t}d_t\left(\Phi_i,\Phi'_i\right)+
\sum\limits_{i=1}^{N_b}d_b\left(\Theta_i,\Theta'_i\right)\right),\\
d_t(\Phi_i,\Phi'_i)&=&
{\rm min} \left(|\Phi_i-\Phi'_i|,2\pi-|\Phi_i-\Phi'_i| \right),\nonumber\\
d_b(\Theta_i,\Theta'_i)&=&|\Theta_i-\Theta'_i|.\nonumber
\end{eqnarray*}
Since $-\pi\le \Phi_i\le \pi$ and $0\le\Theta_i\le\pi$ it follows immediately that
$0\le d_{t,b}\le\pi$. The overlap is unity, if all angles of the conformations
${\bf X}$ and ${\bf X'}$ coincide, else $0\le Q<1$.
\begin{table}[t]
\caption{\label{tabSEQ} Sequences used in the study of thermodynamic
properties of heteropolymers as introduced in Ref.~\cite{irb1}. The number of hydrophobic monomers
is denoted by $\#A$.}
\begin{tabular}{llc}\hline\hline
No. & \multicolumn{1}{c}{sequence} & \#A  \\ \hline
20.1 & BA$_6$BA$_4$BA$_2$BA$_2$B$_2$ & 14\\
20.2 & BA$_2$BA$_4$BABA$_2$BA$_5$B & 14 \\
20.3 & A$_4$B$_2$A$_4$BA$_2$BA$_3$B$_2$A & 14\\
20.4 & A$_4$BA$_2$BABA$_2$B$_2$A$_3$BA$_2$ & 14\\
20.5 & BA$_2$B$_2$A$_3$B$_3$ABABA$_2$BAB & 10\\
20.6 & A$_3$B$_2$AB$_2$ABAB$_2$ABABABA & 10\\ \hline \hline
\end{tabular}
\end{table}

Note that both models are energetically invariant under reflection symmetry. 
This means that also the landscape of the free energy as function of the bond and torsion angles 
is trivially symmetric with respect to the torsional degrees of freedom. Consequently, unless
exceptional cases, there is thus a trivial twofold energetic degeneracy of the global energy minimum,
but the respective rmsd and overlap parameter are different for the associated conformations.
The values quoted throughout the paper were obtained by comparing the lowest-energy conformation
found with both reference conformations
and quoting the value indicating better coincidence (i.e., lower rmsd and higher overlap). This
obvious ambiguity can be circumvented by adding a symmetry-breaking term to the models that 
disfavors, e.g., left-handed helicity~\cite{kaya2}.  

In Table~\ref{tabRMSD} we list the values of both parameters for the lowest-energy conformations
found for the Fibonacci sequences of Table~\ref{tabGEM_A}. We see that for the shortest
sequences (with 13 and 21 monomers) the coincidence of the lowest-energy structures found is
extremely good and we are pretty sure that we have found the ground states. In the case
of the 34mer, modelled with AB model I, both structures also coincide still very well and it seems that the attraction
valley towards the ground state was found within the multicanonical simulation. 
In the simulation of the 34mer with model II the situation is different. As seen from Table~\ref{tabGEM_B},
we found surprisingly the marginally lower energy value in the multicanonical simulation, but the associated
conformation differs significantly from that identified with ELP. It is likely that both conformations do not 
belong to the same attraction basin
and it is a future task to reveal whether this is a first indication of metastability.
The situation is even more complex for the 55mer. As the parameters tell us, the lowest-energy conformations
identified by multicanonical simulation and ELP minimization show significant structural
differences in both models. 

Answering the question of metastability is strongly related with the problem of identifying the folding path or an
appropriate parameterization of the free energy landscape. Due to hidden barriers it is 
practically impossible that ordinary stand-alone multicanonical sampling will be able
to achieve this for such relatively long sequences. Note that, in our model I 
multicanonical simulation for the 55mer, 
we precisely sampled the density of states over {\em 120 orders of magnitude}, i.e., the probability for
finding randomly the lowest-energy conformation (that we identified with multicanonical sampling)
in the conformational space is even less than $10^{-120}$!       
In Fig.~\ref{min55elp} we show two views of the global energy minimum conformation
with energy $E_{\rm min}\approx -42.4$ for the 55mer found by applying the ELP
minimizer to AB model I. The hydrophobic core is tube-like and the chain 
forms a helical structure (which is here an intrinsic property of the model and is not due
to hydrogen-bonding obviously being not supplied by the model). 
\subsection{Thermodynamic properties of heteropolymers}
\label{thermo}
\begin{table}
\caption{\label{tabTHERM1} Minimal energies 
and temperatures of the maximum specific heats for the six 20mers of Table~\ref{tabSEQ} 
using AB model I as obtained by multicanonical sampling. The global maximum of the respective
specific heats is indicated by a star ($\star$). 
For comparison, we have also given the globally minimal energies
found from minimization with ELP as well as the respective rmsd and the 
structural overlap parameter $Q$ of the corresponding minimum energy conformations.}
\begin{tabular}{llllllll}\hline\hline
No. & \multicolumn{1}{c}{$E_{\rm min}^{\rm MUCA}$} & 
\multicolumn{1}{c}{$E_{\rm min}^{\rm ELP}$} & \multicolumn{1}{c}{rmsd} &
\multicolumn{1}{c}{$Q$} & \multicolumn{1}{c}{$T_C^{(1)}$} & \multicolumn{1}{c}{$T_C^{(2)}$}\\ \hline
20.1 & $-33.766$ & $-33.810$ & $0.048$ & $0.954$ & $0.27(3)^\star$ & $0.61(5)$\\
20.2 & $-33.920$ & $-33.926$ & $0.015$ & $0.992$ & $0.26(4)^\star$ & $0.69(4)$\\
20.3 & $-33.582$ & $-33.578$ & $0.025$ & $0.990$ & $0.25(3)^\star$ & $0.69(3)$\\
20.4 & $-34.496$ & $-34.498$ & $0.030$ & $0.985$ & $0.26(3)$ & $0.66(2)^\star$ \\
20.5 & $-19.647$ & $-19.653$ & $0.017$ & $0.988$ & $0.15(2)$ & $0.41(1)^\star$ \\
20.6 & $-19.322$ & $-19.326$ & $0.047$ & $0.989$ & $0.15(2)\footnote{Specific heat of sequence 20.6 
possesses only one maximum at $T_C^{(2)}\approx 0.35$. The value given for $T_C^{(1)}$ belongs to
the pronounced turning point.}$ & $0.35(1)^\star$ \\ \hline \hline
\end{tabular}
\end{table}
\begin{table}
\caption{\label{tabTHERM2} Same as Table~\ref{tabTHERM1}, but using AB model II. For comparison
the specific heat maximum locations $T_s$, estimated in Ref.~\cite{irb1}, are also given.}
\begin{tabular}{llllllll}\hline\hline
No. & \multicolumn{1}{c}{$E_{\rm min}^{\rm MUCA}$} & 
\multicolumn{1}{c}{$E_{\rm min}^{\rm ELP}$} & \multicolumn{1}{c}{rmsd} &
\multicolumn{1}{c}{$Q$} & \multicolumn{1}{c}{$T_C$} & 
\multicolumn{1}{c}{$T_s$~\cite{irb1}}\\ \hline
20.1 & $-58.306$ & $-58.317$ & $0.006$ & $0.999$ & $0.35(4)$ & $0.36$ \\
20.2 & $-58.880$ & $-58.914$ & $0.009$ & $0.997$ & $0.33(4)$ & $0.32$ \\
20.3 & $-59.293$ & $-59.338$ & $0.010$ & $0.997$ & $0.29(3)$ & $0.30$ \\
20.4 & $-59.068$ & $-59.079$ & $0.007$ & $0.998$ & $0.27(4)$ & $0.27$ \\
20.5 & $-51.525$ & $-51.566$ & $0.012$ & $0.998$ & $0.33(5)$ & $0.33$ \\
20.6 & $-53.359$ & $-53.417$ & $0.014$ & $0.996$ & $0.25(2)$ & $0.26$ \\ \hline \hline
\end{tabular}
\end{table}
Our primary interest of this study is focussed on thermodynamic properties of heteropolymers,
in particular on conformational transitions heteropolymers pass from random coils to
native conformations with compact hydrophobic core. We investigated six heteropolymers with
20 monomers as studied by Irb\"ack {\em et al.}\ in Ref.~\cite{irb1} with AB model II. 
The associated sequences are listed in Table~\ref{tabSEQ}.
Notice that the hydrophobicity ($=\#A$ monomers in the sequence) is identical
($=14$) for the first four sequences 20.1--20.4, while sequences 20.5 and 20.6
possess only 10 hydrophobic residues.
In Ref.~\cite{irb1},
the thermodynamic behavior of these sequences was studied by means of the simulated tempering 
method. For revealing low-temperature properties, an additional quenching procedure was
performed. In our simulations, we used multicanonical sampling over the entire range of temperatures
without any additional quenching.
\subsubsection{Multicanonical sampling of heteropolymers with 20 monomers}
\label{sectgs}
\begin{figure}
\centerline{\epsfxsize=8.8cm \epsfbox{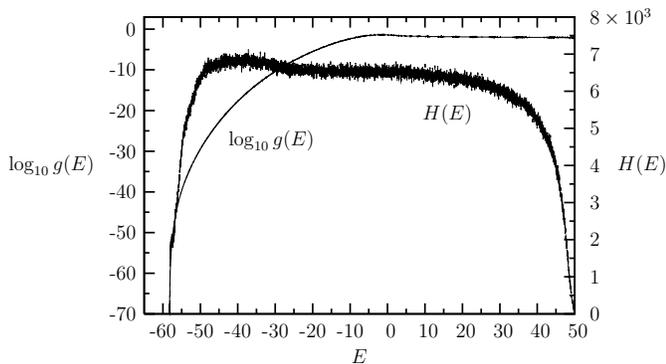}}
\caption{\label{gE20.2} Density of states $g(E)$ (normalized to unity over the
plotted energy interval) and flat multicanonical histogram
$H(E)$ for sequence 20.1 (AB model II).} 
\end{figure}

As described in Sec.~\ref{GEM}, the multicanonical method is capable to find even the lowest-energy states
without any biasing or quenching. We proved this also for the 20mers of Table~\ref{tabSEQ} 
by comparing once more with the 
ELP minimization method. In Tables~\ref{tabTHERM1} and~\ref{tabTHERM2} we present
the estimates of the global energy minima in both models we found in these simulations. Once more,
the values obtained with multicanonical sampling agree pretty well with those from ELP
minimization. The respective structural coincidences are confirmed by the values for the rmsd and the
overlap also being given in these tables.   
In order to identify conformational transitions, we calculated the specific heat 
$C_V(T)=(\langle E^2\rangle-\langle E\rangle^2)/k_BT^2$ with
$\langle E^k\rangle=\sum_E g(E) E^k \exp(-E/k_BT)/\sum_E g(E) \exp(-E/k_BT)$ from the density
of states $g(E)$. The density of states was found (up to an unimportant overall
normalization constant) by reweighting the multicanonical
energy distribution obtained with multicanonical sampling to the canonical distribution 
$P^{{\rm can},T}(E)$ at infinite temperature ($\beta\equiv 1/k_BT = 0$), since
$g(E)=P^{{\rm can},\infty}(E)$. Figure~\ref{gE20.2} shows, as an example, the density of
states $g(E)$ (normalized to unity) and the multicanonical histogram for sequence 20.1 simulated with AB model II. 
We sampled conformations with energy values lying in the interval $[-60.0,50.0]$,
discretized in bins of size $0.01$, and 
required the multicanonical histogram to be flat for at least 70\% of the core of this interval, i.e., within
the energy range $[-43.5,33.5]$. Within and partly beyond this region, we achieved 
almost perfect flatness, i.e., the ratios between the mean and maximal histogram value, 
$H_{\rm mean}/H_{\rm max}$, as well as the ratio between minimum and mean, $H_{\rm min}/H_{\rm mean}$, 
exceeded $0.9$. As a consequence of this high-accurate sampling of the energy space this enabled us to
calculate the density of states very precisely over about 70 orders of magnitude. The energy scale 
is, of course, bounded from below by the ground-state energy $E_{\rm min}$, and that we closely 
approximated this value can be seen by the strong decrease of the logarithm of the density of 
states for the lowest energies. This strong decrease of the density of states curves near
the ground-state energy is common to all short heteropolymer sequences studied. 
It reflects the isolated character of the ground state within the energy landscape.    
\begin{figure*}[t]
\centerline{\epsfxsize=17.6cm \epsfbox{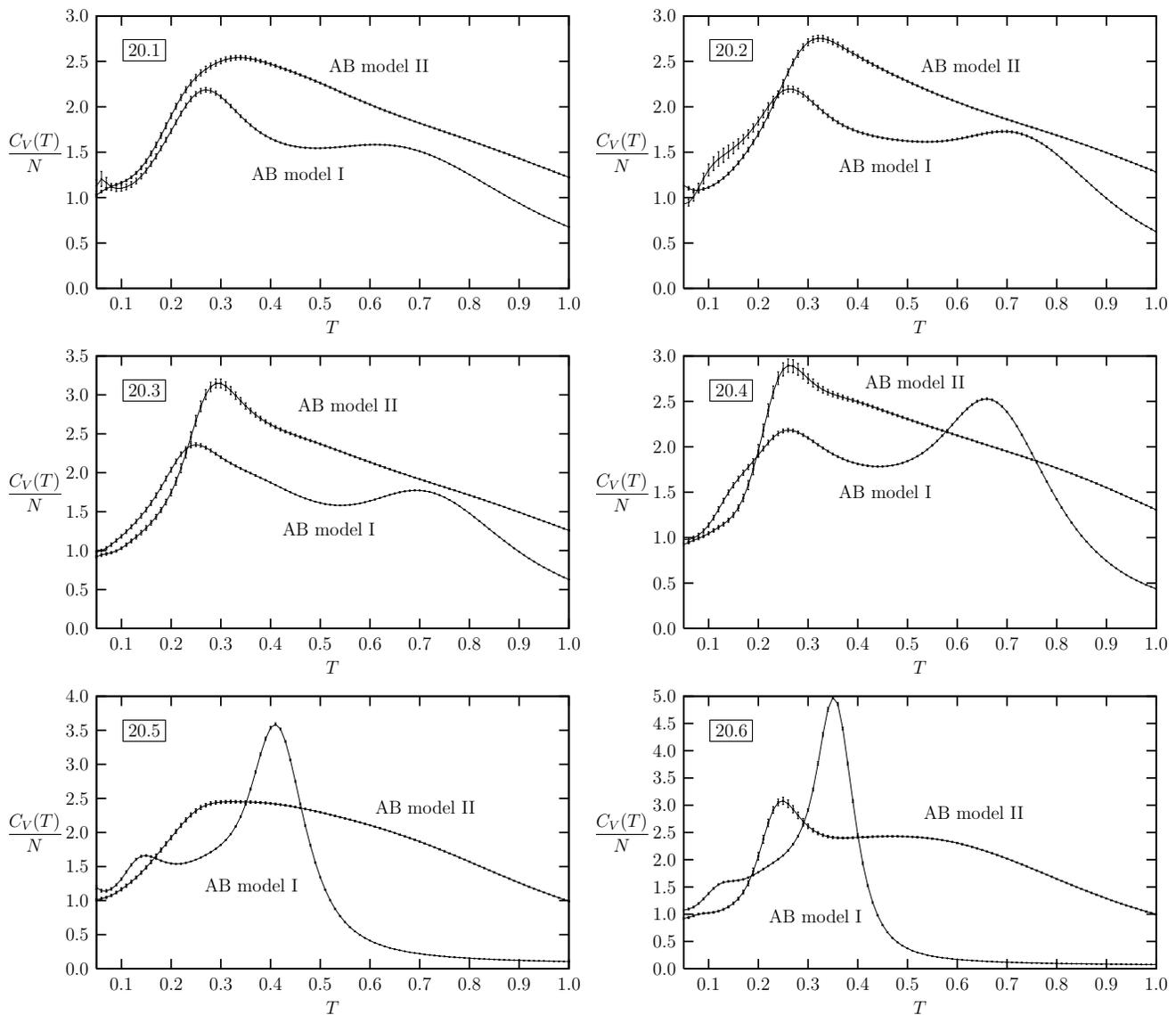}}
\caption{\label{figC20} Specific heats of the 20mers listed in Table~\ref{tabSEQ}.} 
\end{figure*}
\begin{figure}
\centerline{\epsfxsize=8.8cm \epsfbox{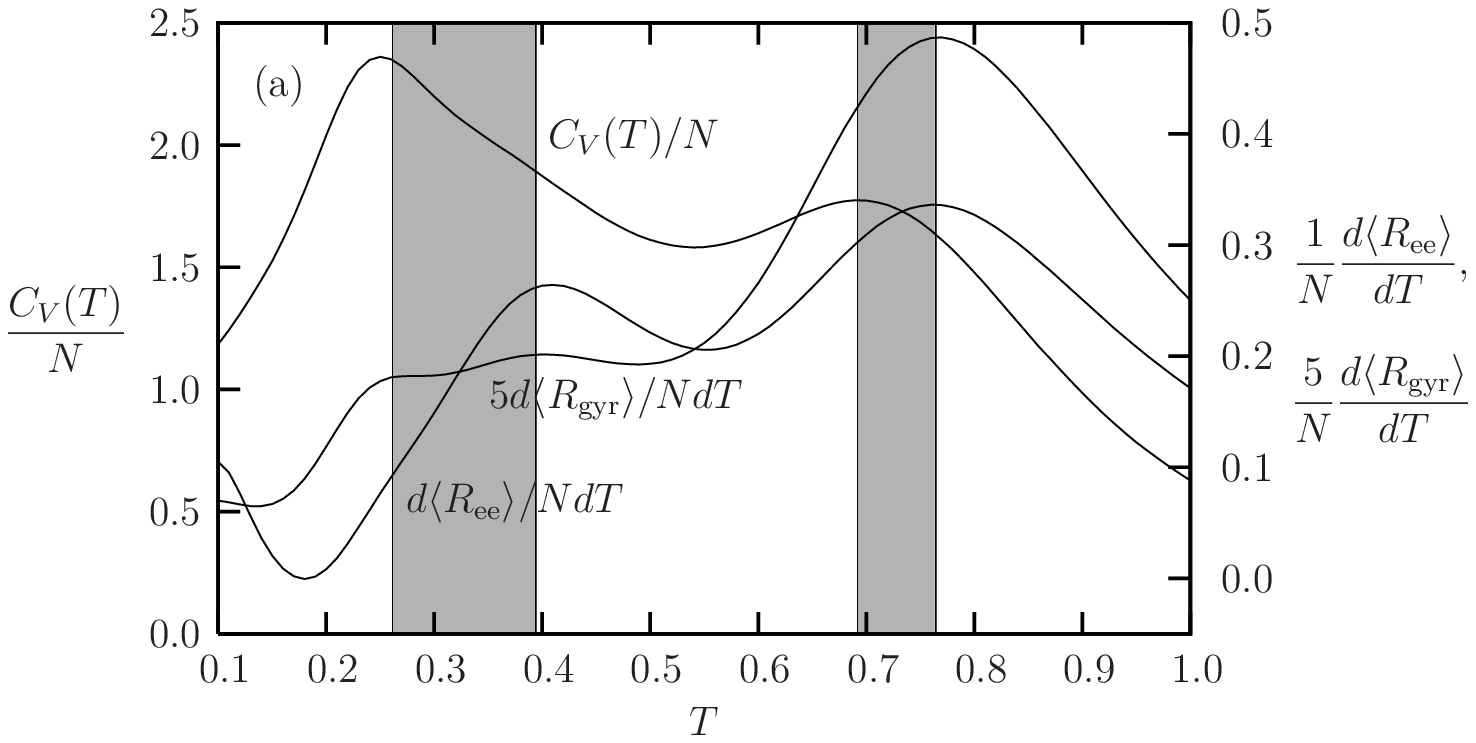}}

\centerline{\epsfxsize=8.8cm \epsfbox{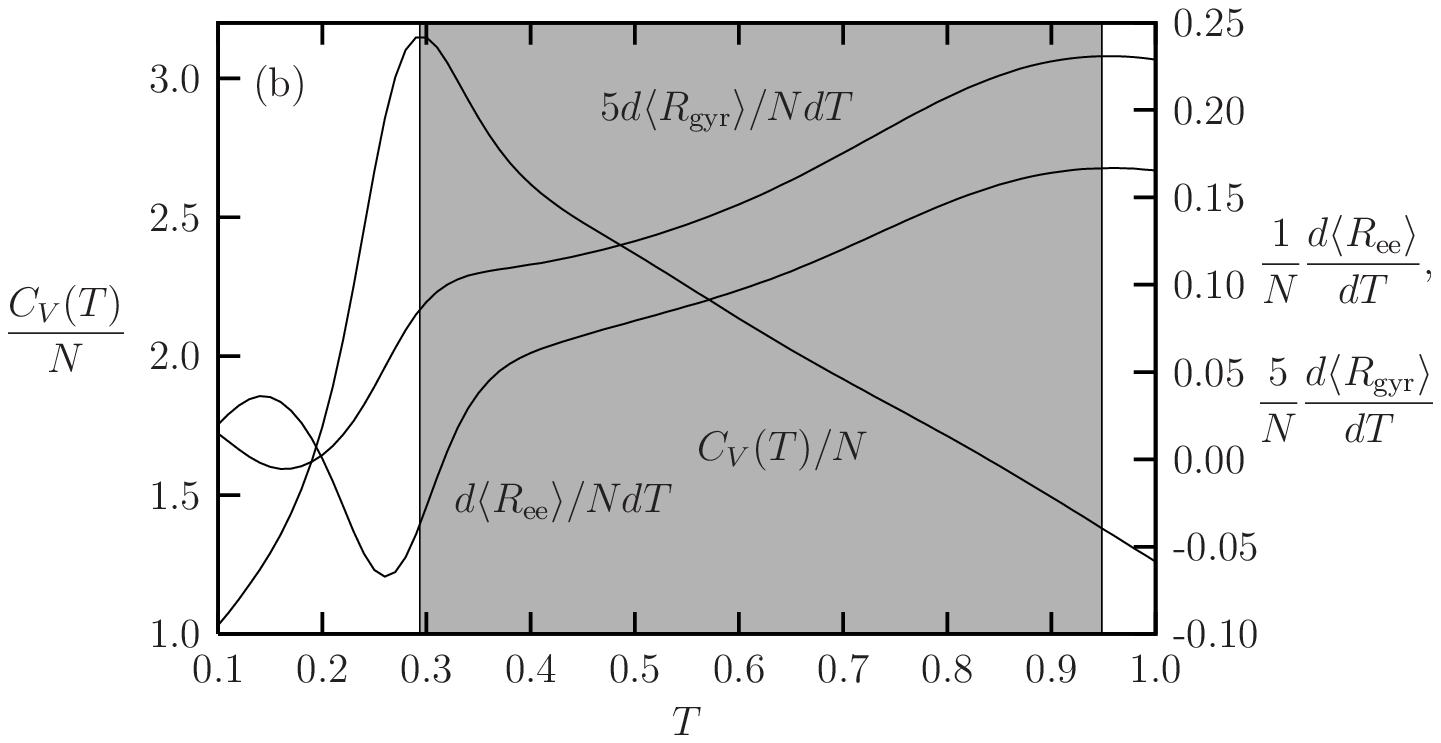}}
\caption{\label{fl20.4} Fluctuations of energy (specific heat), radius of gyration and end-to-end distance 
for sequence 20.3 from simulations with (a) AB model I and (b) AB model II.} 
\end{figure}

We used the density of states to calculate the specific heats of the 20mers
in both models. The results are shown in Fig.~\ref{figC20}. A first observation
is that the specific heats obtained from simulations of AB model I show up
two distinct peaks with the low-temperature peak located at $T_C^{(1)}$ and 
the high-temperature peak at $T_C^{(2)}$ compiled in Table~\ref{tabTHERM1}.
The AB model II, on the other hand, favors only one pronounced peak 
at $T_C$ with a long-range
high-temperature tail. In Table~\ref{tabTHERM2} 
we compare our peak temperatures $T_C$ 
with the values given in Ref.~\cite{irb1}. They are in good correspondence.

The sequences considered here are very short and the 
native fold contains a single hydrophobic core.  
Interpreting the curves for the specific heats in Fig.~\ref{figC20} in terms of
conformational transitions, we conclude that
the heteropolymers simulated with AB model I tend to form, within the temperature region 
$T_C^{(1)}<T<T_C^{(2)}$, intermediate states (often
also called traps) comparable with globules in the collapsed phase of polymers.
For sequences 20.5 and, in particular, 20.6 the smaller number of hydrophobic monomers
causes a much sharper transition at $T_C^{(2)}$ than at $T_C^{(1)}$ (where, in fact,
the specific heat of sequence 20.6 possesses only a turning point). The pronounced transition 
near $T_C^{(2)}$ is
connected with a dramatic change of the radius of gyration, as can be seen later in Fig.~\ref{figRgyr20},
indicating the collapse from stretched to highly compact conformations with decreasing temperature.  
The conformations dominant for high temperatures $T>T_C^{(2)}$ are random coils, while for
temperatures $T<T_C^{(1)}$ primarily conformations with compact hydrophobic core are favored.
The intermediary globular ``phase'' is not at all present for the exemplified sequences 
when modeled with AB model II, where
only the latter two ``phases'' can be distinguished. We have to remark that
what we denote ``phases'' are not phases in the strict thermodynamic sense, since 
for heteropolymers of the type we used in this study (this means we are not focussed on
sequences that have special symmetries, as for example diblock copolymers A$_n$B$_m$),
a thermodynamic limit is {\em in principle} nonsensical. Therefore conformational
transitions of heteropolymers are not true phase transitions. As a consequence, fluctuating
quantities, for example the derivatives with respect to the temperature of the mean 
radius of gyration, $d\langle R_{\rm gyr}\rangle/dT$, and the mean end-to-end distance, 
$d\langle R_{\rm ee}\rangle/dT$, do not indicate conformational activity at the same
temperatures, as well as when compared with the specific heat. 

We conclude that conformational
transitions of heteropolymers happen within a certain interval of temperatures, not at a fixed critical
temperature. This is a typical finite-size effect and, for this reason, the peak temperatures
$T_C^{(1)}$ and $T_C^{(2)}$ (for model I) and $T_C$ (for model II) defined above for the
specific heat are only representatives for the entire intervals.
In order to make this more explicit, we consider sequence 20.3 in more detail. In Figs.~\ref{fl20.4}(a) 
(for AB model I) and \ref{fl20.4}(b) (AB model II), we compare the 
energetic fluctuations (in form of the specific heat $C_V$) with the respective fluctuations of radius 
of gyration and end-to-end distance, 
$d\langle R_{\rm ee,gyr}\rangle/dT=(\langle R_{\rm ee,gyr}E\rangle-\langle R_{\rm ee,gyr}\rangle\langle E\rangle)/k_BT^2$.
Obviously, the temperatures with maximal fluctuations
are not identical and the shaded areas are spanned over the temperature intervals, 
where strongest activity is expected. 
We observe for this example that in model I
(Fig.~\ref{fl20.4}(a)) two such centers of activity can be separated linked by an
intermediary interval of globular traps. In fact,
there is a minimum of the specific heat at $T_C^{\rm min}$ between $T_C^{(1)}$
and $T_C^{(2)}$, but the increase of internal energy by forming these states, which is given by
$\int_{T_C^{(2)}}^{T_C^{\rm min}}dT\,C_V(T)$, is rather small and the 
globules are not very stable. From Fig.~\ref{fl20.4}(b), we conclude for this sequence that in
model II no peculiar intermediary conformations occur, and the folding of the heteropolymer 
is a one-step process.

It is widely believed and experimentally consolidated that
realistic short single-domain proteins are usually two-state folders~\cite{chan1}. This means,
there is only one folding transition and the protein is either in the folded or an
unfolded (or denatured) state. Therefore, AB model II could indeed serve as a simple
effective model for two-state heteropolymers. The main difference between both
models under study is that model II contains an implicit torsional energy which
is not present in model I. This is in correspondence with more knowledge-based
G\={o}-like models with explicit torsional energy contributions for the study of small 
proteins with known typical two-state folding--unfolding kinetics~\cite{kaya1}. 

Nonetheless, there are also examples of small peptides exhibiting two clear peaks
in the specific heat. In Ref.~\cite{hansmann2}, the artificial peptide Ala$_{10}$Gly$_5$Ala$_{10}$
was studied in detail and it turned out that two transitions separate the ground-state
conformation and random coil states. One is the alanine mediated helix-coil transition
and the second the formation of a glycine hairpin that leads to a more compact conformation. 
\subsubsection{Comparison with the homopolymer}
\label{secthom}
\begin{figure}
\centerline{\epsfxsize=8.8cm \epsfbox{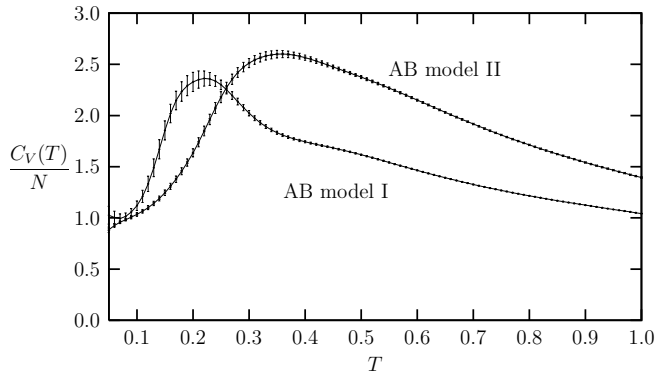}}
\caption{\label{figAAC20} Specific heats of the homopolymer $A_{20}$ with 20 monomers for both models.}
\end{figure}
It is also interesting to compare the thermodynamic behavior of the heteropolymers with 20 monomers
as described in the previous section with the homopolymer consisting of 20 $A$-type monomers, $A_{20}$.
This is the consequent off-lattice generalisation of self-avoiding interacting walks on the lattice (ISAW)
that have been extensively studied over the past decades. In contrast to heteropolymers, where, because of
the associated sequence of finite length, a thermodynamic limit does not exist, homopolymers show up
a characteristic second-order phase transition between random coil conformations (``good solvent'') and compact
globules (``poor solvent''), the so-called $\Theta$ transition~\cite{deGennes1}. 

In this study, our interest was not focussed on the $\Theta$ transition but more on the direct comparison
of the finite-size homopolymer and the different heteropolymer sequences. In Fig.~\ref{figAAC20} we
have plotted the specific heats of this homopolymer for both models under study. The first observation is
that, independent of the model, the collapse from random coils (high temperatures) to globular 
conformations (low temperatures) happens, roughly, in one step. There is only one energetic
barrier as indicated by the single peak of the specific heats. 
\begin{figure}
\centerline{\epsfxsize=8.8cm \epsfbox{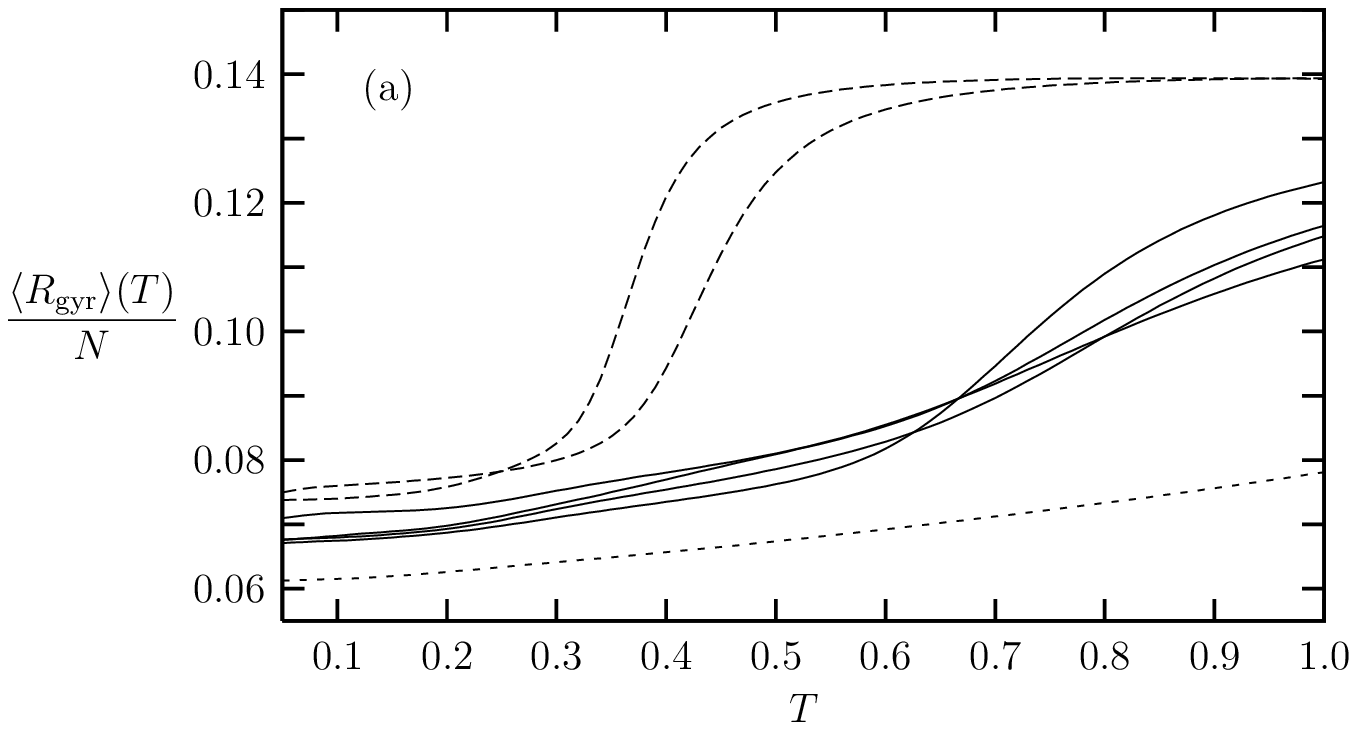}}

\centerline{\epsfxsize=8.8cm \epsfbox{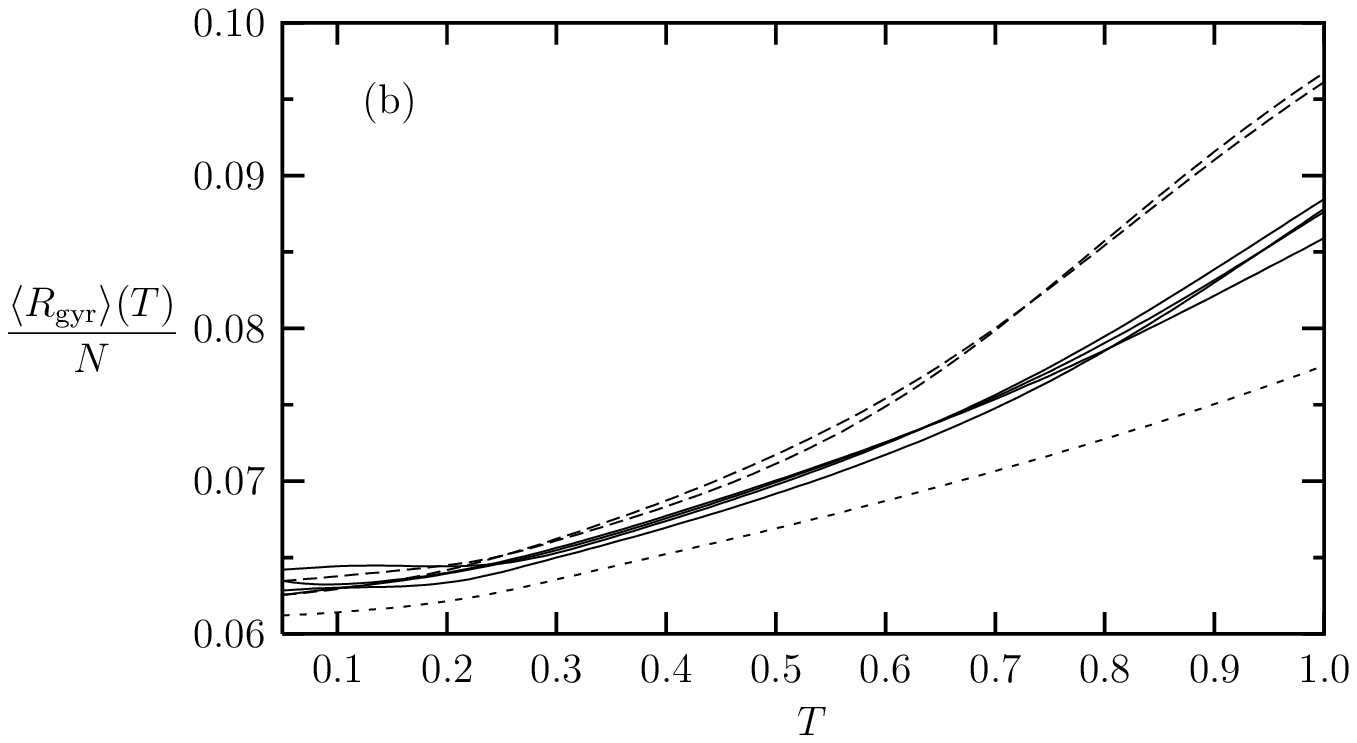}}
\caption{\label{figRgyr20} Mean radius of gyration $\langle R_{\rm gyr}\rangle$ as a function
of the temperature $T$ for the sequences 20.1--20.4 (solid curves), 20.5, 20.6 (long dashes)
and the homopolymer $A_{20}$ (short-dashed curve) for (a) AB model I and (b) AB model II.} 
\end{figure}

In Fig.~\ref{figRgyr20} we have plotted for both models the mean radii of gyration as a function of 
the temperature for the sequences from Table~\ref{tabSEQ} in comparison with the homopolymer. 
For all temperatures in the interval plotted, the homopolymer obviously takes more compact 
conformations than the heteropolymers, since its mean radius of gyration is always smaller.
This different behavior is an indication for a 
rearrangement of the monomers that is particular for heteropolymers: the formation of 
the hydrophobic core surrounded by the hydrophilic monomers. Since the homopolymer trivially also takes
in the ground state a hydrophobic core conformation (since it only consists of hydrophobic monomers),  
which is obviously more compact than the complete conformations of the heteropolymers, we conclude
that hydrophobic monomers weaken the compactness of low-temperature conformations. Thus,
homopolymers and heteropolymers show a different ``phase'' behavior in the dense phase.
Homopolymers fold into globular conformations which are hydrophobic cores with maximum number
of hydrophobic contacts. Heteropolymers also form very compact hydrophobic cores which are,
of course, smaller than that of the homopolymer due to the smaller number of hydrophobic monomers
in the sequence. In total, however, heteropolymers are less compact than homopolymers because
the hydrophilic monomers are pushed off the core and arrange themselves in a shell around the 
hydrophobic core. For model I, we also see in Fig.~\ref{figRgyr20}(a) a clear tendency
that the mean radius of gyration and thus the compactness strongly depends on the hydrophobicity
of the sequence, i.e., the number of hydrophobic monomers. The curves for 
sequences 20.5 and 20.6 (long-dashed
curves) with 10 $A$'s in the sequence can clearly be separated from the other heteropolymers
in the study (with 14 hydrophobic monomers) and the homopolymer. This supports the
assumption that for heteropolymers the formation of a hydrophobic core is more
favorable than the folding into an entire maximally compact conformation. 
\section{Summary}
\label{summary}
We have investigated two coarse-grained off-lattice heteropolymer models of AB type
that mainly differ in the modelling of energetic bending and torsional barriers. While the original
AB model (model I)~\cite{ab1}, which was first introduced for two-dimensional heteropolymers, treats 
the polymer as a stiff chain of hydrophobic ($A$) and hydrophilic ($B$) monomers without considering
torsional barriers, model II favors bending and an additional contribution containing also 
torsional energy is regarded~\cite{irb1}.
Another noticeable property of model I is that contacts between $A$ and $B$ monomers are
always suppressed, in contrast to model II. 

We studied these models by means of the multicanonical Monte Carlo method. In a first test, we checked
the ability of the algorithm to find lowest-energy conformations. The results were compared
with minimum energy values obtained with the energy-landscape paving (ELP) algorithm by measuring
the rmsd and a generalized overlap parameter that in addition to torsional degrees
of freedom also allows the comparison of bond angles. We found 
very good coincidences for minimum energies and associated conformations for all 
sequences under study, with the exception of the 34mer in model II and the 55mer in both models. 
In the latter case the
random walk in the energy space, which is considered as the system parameter, in not
sufficient to find the global energy minimum and a more detailed study of the
origin of the free energy barriers, i.e., the identification of an appropriate 
order parameter, is required. Nonetheless, for all sequences we obtained much lower values for the 
respective putative global-energy minimum than formerly quoted in the literature using an off-lattice
chain-growth algorithm~\cite{hsu2} and also considerably lower values than with the
annealing contour Monte Carlo (ACMC) method~\cite{liang1}.

Our main objective was the comparison of the conformational transitions in both models. We primarily
studied energetic and conformational fluctuations of several sequences with 20 monomers
and found that in model I there is a general
tendency that, independent of the sequence, the folding from random coils (high temperatures)
to lowest-energy conformations is a two-step process and the folding is slowed down 
by weakly stable intermediary conformations. This is different in model II, were
traps are avoided and the heteropolymers exhibit a two-state folding behavior. 
This is (qualitatively!) in correspondence with the observation that many short
peptides seem to possess a rather smooth free-energy landscape, where only a single barrier
separates unfolded states and native fold~\cite{chan1}. Most of the previous
investigations in this regard were performed by means of a kind of knowledge-based
potentials, where topological properties (e.g., the native contacts) of the folded
state explicitly enter. These potentials allow then quantitative studies of the dynamics of 
the folding process for the specified protein. In our study, however, we were more interested
in the influence of basic principles on the folding transition and therefore 
quantitative comparison with the folding of realistic proteins is not appropriate
at this level of abstraction.    

In order to reveal folding properties specific to heteropolymers, we also
compared with a purely hydrophobic polymer. Our results for the homopolymer
are in accordance with the widely accepted view of the formation of compact
globular conformations below the $\Theta$ transition temperature. The 
globules, which are in our interpretation only  
compact hydrophobic cores with maximum number of hydrophobic contacts,
minimize
the surface exposed to the (implicit) solvent. This behavior was observed
for both models which differ, for the purely hydrophobic homopolymer, only in
local covalent bond properties, but not in 
the non-bonded interaction between the monomers. Since heteropolymers with the same
chain length contain less hydrophobic monomers, the hydrophobic core 
is, although still very dense, smaller and the hydrophilic monomers
form a shell surrounding the core and screening it from the solvent. This,
in consequence, leads to a native conformation that qualitatively differs from 
the typical globules known from polymers. In models allowing for intermediary
states (such as model I in our study), the globular ``phase'' is actually present
for heteropolymers, too, and is dominant in a temperature interval
between the hydrophobic-core phase at very low temperatures and the
random coils that are present at high temperatures.        
  
In conclusion, simple, coarse-grained off-lattice heteropolymer 
models without specific or knowledge-based parameterization are attractive
as they allow for the study of
how basic energetic contributions, bonded and non-bonded, in the model are related to and 
influence the conformational folding process of heteropolymers, respectively.
Herein, the main focus is on the study of general properties
of these systems. In perspective, such models, after refinements with respect
to a few specific interatomic interactions (e.g., hydrogen bonds) will be
suitable for studying heteropolymers of a few hundreds of monomers even quantitatively
without the requirement of a prior knowledge of the final fold.  
\section{Acknowledgements}
M.B.\ thanks Thomas Weikl for interesting discussions on the folding behavior
of short peptides. H.A.\ acknowledges support by the exchange programme of the 
Deutsche Forschungsgemeinschaft (DFG) and T\"UBITAK under contract Nos.\ 446 T\"UR 
112/7/03 and 112/14/04.
This work is partially supported by the German-Israel-Foundation (GIF) under
contract No.\ I-653-181.14/1999. 
\end{document}